\documentclass[12pt,a4j]{article}
\usepackage[dvips]{graphicx}
\begin{document}
\baselineskip=7mm
\centerline{\bf @The Lyapunov stability of the N-soliton solutions  }\par
\centerline{\bf  in the Lax hierarchy of the Benjamin-Ono equation} \par
\bigskip
\centerline{Yoshimasa Matsuno\footnote{{\it E-mail address}: matsuno@yamaguchi-u.ac.jp}}\par
\centerline{\it Division of Applied  Mathematical  Science,} \par
\centerline{\it  Graduate School of Science and
Engineering,} \par
\centerline{\it Yamaguchi University, Ube 755-8611, Japan} \par
\bigskip

\noindent{\bf  Abstract}\par
The Lyapunov stability is established for the $N$-soliton solutions in the Lax hierarchy of the Benjamin-Ono (BO)
equation. We characterize the $N$-soliton profiles  as critical points of certain
Lyapunov functional. By using several results
derived by the inverse scattering transform of the BO equation, 
we  demonstrate the convexity of the Lyapunov functional when evaluated at the $N$-soliton profiles.
From this fact, we deduce that  the N-soliton solutions are energetically stable.

\newpage
\leftline{\bf  I. INTRODUCTION} \par
The Benjamin-Ono (BO) equation describes the unidirectional propagation 
of long internal waves in
stratified fluids of great depth. It may be written in an appropriate dimensionless form as 
$$u_t+2uu_x+Hu_{xx}=0. \eqno(1.1a)$$
Here, $u=u(x,t)$ represents the amplitude of wave, $H$ is the Hilbert transform given by
$$Hu(x,t)={1\over \pi} P\int_{-\infty}^{\infty}{u(y,t)\over y-x}dy, \eqno(1.1b)$$
and the subscripts $t$ and $x$ appended to $u$ denote partial differentiation. The BO equation
can be written as an infinite-dimensional completely integrable Hamiltonian 
dynamical system.$^{1,2}$  A common
feature of  integrable evolution equations is the existence of an infinite sequence of conservation
laws.  The Lax hierarchy of the BO equation is generated by the conservation laws, which
we shall denote by $I_{n} (n= 2, 3, 4 ...)$.  
The first three of $I_n$ read
$$I_2={1\over 2}\int^\infty_{-\infty}u^2dx, \eqno(1.2a)$$
$$I_3=-\int^\infty_{-\infty}\left({1\over 3}u^3+{1\over 2}uHu_x \right)dx, \eqno(1.2b)$$
$$I_4=\int^\infty_{-\infty}\left({1\over 4}u^4+{3\over 4}u^2Hu_x+{1\over 2}u_x^2\right)dx.
\eqno(1.2c)$$
In (1.2), the mass conservation has been excluded since it is irrelevant in the   
present analysis.
The BO hierarchy is defined by the following
nonlinear evolution equations
$${\partial u\over \partial t_n}={\partial\over \partial x }{\delta I_{n+2}\over \delta u},
(n=0, 1, 2, ...), \eqno(1.3)$$
where $\delta/\delta u$ is the variational derivative defined by
$${\partial\over\partial\epsilon}I_{n+2}(u+\epsilon v)|_{\epsilon=0}
=\int^\infty_{-\infty}{\delta I_{n+2}\over \delta u(x)}v(x)dx.
\eqno(1.4)$$
When $n=1$,  (1.3) becomes the BO equation (1.1) with the identification $t_1=t$ 
while when $n=2$, it yields the first  higher-order BO equation$^{3}$
$$u_{t_2}=\left(u^3+{3\over 2}uHu_x+{3\over 2}H(uu_x)-u_{xx}\right)_x.
\eqno(1.5)$$
Note that the first member of (1.3) reduces simply to a linear equation $u_{t_0}=u_x$.
As will be shown below, all  the members of the BO hierarchy exhibit the $N$-soliton solution
characterized by the $2N$ parameters  $a_j$ and $x_{j0} (j=1, 2, ..., N)$
where $N$ is an arbitrary positive integer:
$$u=u_N(x-x_1, x-x_2, ..., x-x_N). \eqno(1.6a)$$
Here
$$x_j=\sum_{s=0}^\infty (-1)^{s+1}{s+1\over 2^s}a_j^{s}t_s+x_{j0}, 
(j=1, 2, .., N), \eqno(1.6b)$$
 $a_j$ are amplitude parameters  satisfying the conditions $a_j>0, a_j\not=a_k$ for
$j\not=k (j, k=1, 2, ..., N)$ and $x_{j0}$ are arbitrary phase parameters. 
Explicitly, $u_N$ has a simple expression in terms of a tau function $f$
$$u_N=i{\partial\over \partial x}{\rm ln}\ {f\over f^*}, \  f={\rm det}\ F, 
\eqno(1.7a)$$
where $F=(f_{jk})_{1\leq j,k\leq N}$ is an $N\times N$ matrix with elements
$$ f_{jk}=\left(x-x_j+{i\over a_j}\right)\delta_{jk}
-{2i\over a_j-a_k}(1-\delta_{jk}). \eqno(1.7b)$$
Here, $f^*$ is a complex conjugate of $f$ and $\delta_{jk}$ is Kronecker's delta.
In particular, for $N=1$, (1.7) represents the 1-soliton solution
with a Lorenzian profile
$$u_1={2a_1\over a_1^2(x-x_1)^2+1}. \eqno(1.8)$$
A direct proof of (1.7) using an elementary theory of determinants will
be presented in Appendix A.@\par
The definition of the stability of solitons may be classified according to the following three 
categories: i) linear (or spectral) stability, ii) energetic stability, iii) nonlinear stability.
The  energetic stability implies that the second variation of certain Lyapunov functional
becomes strictly positive when evaluated at the soliton solutions. It would also lead to the
linear stability since the second variation is preserved for the linearized equation. 
In order to  extend
the energetic stability to the nonlinear stability which deals with small but 
finite amplitude perturbations, one must take into account higher-order
nonlinear terms neglected in evaluating the Lyapunov functional and
 this makes the analysis more difficult.  In  accordance with the above classification of the
stability, we shall briefly review some known results associated with the stability
characteristics of the BO solitons. 
The linear stability of the BO 1-soliton solution has been proved by
solving the eigenvalue problem associated with the linearized BO equation.$^{4}$ 
A subsequent nonlinear analysis shows that the soliton is also stable against small
but finite perturbations.$^{5}$  As for the general $N$-soliton solution, 
its linear stability characteristic has been established
by solving explicitly the initial value problem of the linearlized BO equation
and investigating the large-time asymptotic of the solution.$^{6,7}$ In the process,
the completeness relation for the eigenfunctions of the BO equation
linearized around the $N$-soliton solution has played a central role.
The recent study demonstrates the orbital stability of the 2-soliton solution 
in which the  stability problem has been settled based on the
Lyapunov method combined with the spectral analysis of the 
operators associated with the
linearized BO equation.$^{8}$ 
The approach used in this paper originates from the stability analysis of the
multisoliton solutions of the Korteweg-de Vries (KdV) equation by means of
the constrained variational principle.$^{9}$ See also an analogous work
dealing with the spectral stability of the multisoliton solutions in the
KdV hierarchy.$^{10}$  
All the works mentioned above are concerned with the 
stability of solitons for the BO equation. The stability characteristics of
solitons in the BO hierarchy have not been considered as yet. \par
The purpose of this paper is to establish the Lyapunov stability of the 
general $N$-soliton 
solution (1.6).  To be more specific, let us consider the following higher-order BO equation
which consists of the commuting flows of the BO hierarchy
$$u_t={\partial \over \partial x}{\delta H_{N}\over \delta u}, \eqno(1.9a)$$
where the Lyapunov functional $H_N$ is given by
$$H_N(u)=I_{N+2}+\sum_{n=1}^N\mu_nI_{n+1}, \eqno(1.9b)$$
and $\mu_n$ are  Lagrange multipliers which will be expressed in terms of the elementary 
symmetric functions   of $a_1, a_2, ..., a_N$. See Sec. III
for the detail.
 We define the profile (or shape) of 
the $N$-soliton solution by $U_N=U_N(x)\equiv u_N|_{t_0=t_1=...=0}$.
We observe from (1.7) that the $N$-soliton profile has the
same functional form for all the members of the hierarchy, the
only difference being the velocities of the solitons.
We show that $U_N$ is a stationary solution of (1.9) decaying
at infinity. Namely, $U_N$ is realized as a critical point of the
functional $H_N$.
Using (1.9b), this condition can be written as the Euler-Lagrange equation
$${\delta I_{N+2}\over \delta u}+\sum_{n=1}^N\mu_n{\delta I_{n+1}\over \delta u}=0, \ {\rm at}\ u=U_N.
\eqno(1.10)$$
The Lyapunov stability of $U_N$ may characterize $U_N$ as a minimal point of 
the functional $I_{N+2}$ 
subjected to $N$ constraints
$$I_{n+1}(u)=d_n, (n=1, 2, ..., N), \eqno(1.11)$$
where $d_n$ are real constants and consequently the second variation of $H_N$ is 
strictly positive  at $U_N$. This means that $H_N$ is convex at $U_N$, so that
the following inequality holds
$$ H_N(U_N+\epsilon v)-H_N(U_N)>0, \eqno(1.12)$$
where $\epsilon v$ is a perturbation imposed on $U_N$ which belongs to
certain function space specified by the
$2N$ integral conditions.  We assume that
the $L^2$ norm of  $v$ is finite. The small parameter $\epsilon$
has been introduced to measure the magnitude of the perturbation.
The inequality (1.12) shows that the $N$-soliton solutions are energetically stable.
 We give a direct proof of  (1.12) 
 with the aid of the results obtained by the inverse scattering transform (IST) for
the BO equation.$^{1,2,11,12}$ In Sec. II, we summarize the background results arising
from the perturbation theory and the Hamiltonian formulation of the BO equation  
which provide the necessary machinery in carrying out the stability analysis. In Sec. III, we
prove the inequality and hence establish the Lyapunov stability of the $N$-soliton
solutions in the Lax hierarchy of the BO equation.  
 In Appendix A, we present a direct proof of the $N$-soliton
solution (1.7). In Appendix B, we evaluate the number of positive eigenvalues of
the Hessian matrix associated with $H_N$. \par
\bigskip
\leftline{\bf II. BACKGROUND RESULTS OF IST} \par
The IST has been applied successfully to solve the initial value
problem of the BO equation.$^{11,12}$ Furthermore, for real generic
potentials it has been used to prove  the complete integrability of the
BO equation.$^{1,2}$ Here, we summarize some background results 
of the IST necessary for the stability analysis. \par
\leftline{\bf A. Eigenvalue problem} \par
The eigenvalue problem associated with the IST of the BO equation
may take the form
$$i\phi_x^++\lambda(\phi^+-\phi^-)=-u\phi^+, \eqno(2.1)$$
where $\phi^+(\phi^-)$ is the boundary value of the analytic
function in the upper(lower)-half complex $x$ plane, $u$ is a
real potential rapidly decreasing at infinity and $\lambda$ is the
eigenvalue (or the spectral parameter). We define the two Jost
solutions of (2.1) specified by the boundary condition as
$x\rightarrow +\infty$
$$N(x,\lambda) \rightarrow e^{i\lambda x}, 
\ \bar N(x,\lambda) \rightarrow 1, \eqno(2.2)$$
and the analogous ones as $x\rightarrow -\infty$
$$\bar M(x,\lambda) \rightarrow 1, 
\  M(x,\lambda) \rightarrow e^{i\lambda x}. \eqno(2.3)$$
These solutions satisfy the linear integral equations
$$N_x-i\lambda N=iP_+(uN), \eqno(2.4a)$$
$$\bar N_x-i\lambda \bar N=iP_+(u\bar N)-i\lambda, \eqno(2.4b)$$
$$M_x-i\lambda M=iP_+(u M)-i\lambda, \eqno(2.4c)$$
$$\bar M_x-i\lambda \bar M=iP_+(u\bar M), \eqno(2.4d)$$
where $P_+$ is the projection operator defined by
$P_+={1\over 2}(1-iH)$. The solutions of (2.4) subjected to
the boundary conditions (2.2) and (2.3) exist for $\lambda>0$.
The Jost functions $M, N$ and $\bar N$ are then related by
$$M=\bar N+\beta N, \eqno(2.5)$$
where $\beta$ is a reflection coefficient. For pure soliton
potentials, this reflection coefficient vanishes identically. \par
There exists a set of solutions $\Phi_j(x)$ for negative 
$\lambda=\lambda_j (j=1, 2, ..., N)$ which satisfy the equation
$$\Phi_{j,x}-i\lambda_j\Phi_j=iP_+(u\Phi_j), \ (j=1, 2, ..., N),
\eqno(2.6a)$$
with the boundary conditions
$$\Phi_j \rightarrow {1\over x}, \ x \rightarrow +\infty, 
\ (j=1, 2, ..., N). \eqno(2.6b)$$
\leftline{\bf B. Conservation laws} \par
It follows from (1.1), (2.4b) and the time evolution equation
for $\bar N$ 
$$\bar N_t-2\lambda \bar N_{x}-i\bar N_{xx}-2(P_+u_x)\bar N =0, \eqno(2.7)$$
that the quantity $\int^\infty_{-\infty}u(x,t)\bar N(x,t)dx$
is conserved in time. Expanding $\bar N$ in inverse powers of $\lambda$
$$\bar N=\sum_{n=0}^\infty{(-1)^n\bar N_{n+1}\over \lambda^n}, \ \bar N_1=1,
\eqno(2.8a)$$
and substituting (2.8a) into (2.4b), we obtain the following recursion relation
that determines $\bar N_n$:
$$\bar N_{n+1}=i\bar N_{n,x}+P_+(u\bar N_n), \ n\geq 1. \eqno(2.8b)$$
The $n$th conservation law may be taken as
$$I_n=(-1)^n\int^\infty_{-\infty}u\bar N_ndx, \eqno(2.9)$$
where a factor $(-1)^n$ is multiplied for convenience. 
The first three of $I_n$ except $I_1$ are already given by (1.2). In terms of the
scattering data $\beta$ and $\lambda_j$, $I_n$ can be evaluated as
$$I_n=(-1)^n\left\{2\pi\sum_{j=1}^N(-\lambda_j)^{n-1}+{(-1)^n\over 2\pi}
\int^\infty_0\lambda^{n-2}\beta^*(\lambda)\beta(\lambda)d\lambda\right\},
\ (n=1, 2, ...).\eqno(2.10)$$
The first term on the right-hand side of (2.10) is the contribution
from solitons and the second term comes from radiations. It is important
that both contributions are additive. 
A remarkable feature of the conservation laws is that they are in involution,
namely $I_n (n=1, 2, ...)$ commute each other in an appropriate Poisson bracket.
In particular
$$\int^\infty_{-\infty}\left({\delta I_n\over \delta u(x)}\right)_{u=U_N}
{\partial\over\partial x}\left({\delta I_m\over \delta u(x)}\right)_{u=U_N}dx=0,
\ (n, m=1, 2, ...). \eqno(2.11)$$
\par
\leftline{\bf C. Variational derivatives} \par
The variational derivatives of the scattering date with respect to the
potential are calculated explicitly. In developing the Lyapunov stability, 
we need the formulas of  the variational derivatives evaluated for
the $N$-soliton potential $u=U_N$. In particular, the following formula plays an important role in
our analysis
$$\left({\delta \lambda_j\over \delta u(x)}\right)_{u=U_N}={1\over 2\pi\lambda_j}\Phi_j^*(x)\Phi_j(x),
\ (j=1, 2, ..., N). \eqno(2.12)$$
Here, the eigenfunction $\Phi_j$  corresponding to the discrete spectrum $\lambda_j$
satisfies the system of linear algebraic equations
$$(x-\gamma_j)\Phi_j+i\sum_{k=1\atop (k\not=j)}^N{1\over \lambda_j-\lambda_k}\Phi_k=1,
\ (j=1, 2, ..., N), \eqno(2.13)$$
where $\gamma_j=x_{j0}+i/(2\lambda_j)$ and $x_{j0}$ are real constants.
Recall that  $\lambda_j$  are related to the amplitude parameters $a_j$ introduced
in (1.7) by the relations $\lambda_j=-a_j/2 (j=1, 2, ..., N)$.
Taking account of the fact that the reflection coefficient $\beta$ becomes zero for $u=U_N$,
we can derive from (2.10) and (2.12) the formula
$$\left({\delta I_n\over \delta u(x)}\right)_{u=U_N}=(-1)^n(n-1)
\sum_{j=1}^N(-\lambda_j)^{n-3}\Phi_j^*(x)\Phi_j(x),
\ (n= 2, 3, ..., N). \eqno(2.14)$$
In terms of $\Phi_j$, $U_N$ has the following two alternative expressions:
$$U_N=i\sum_{j=1}^N(\Phi_j-\Phi_j^*), \eqno(2.15)$$
$$U_N=-\sum_{j=1}^N{1\over \lambda_j}\Phi_j^*\Phi_j. \eqno(2.16)$$
The positive definiteness of $U_N$ is obvious from (2.16) since all $\lambda_j$ are
negative quantities. One can derive (2.16) by using (2.13) and (2.15).
The formula (2.16) also follows from (1.2a), (2.10) and (2.14).
In Appendix A, we show that $U_N$ can be rewritten in a compact form in
terms of a determinant. \par
The following relation concerning the  variational derivative of $\beta$ with respect to $u$ 
is useful in evaluating the contribution of the continuous part to the functional $H_N$:
$${\delta\beta(\lambda)\over\delta u(x)}=iM(x,\lambda)N^*(x,\lambda). \eqno(2.17)$$
For the $N$-soliton potential $u=U_N$, $M$ reduces to $\bar N $ by (2.5) and $\beta\equiv 0$.
The  function $MN^*$ satisfies the orthogonality conditions 
$$\int^\infty_{-\infty}M(x,\lambda)N^*(x,\lambda){\partial\over\partial x}
\left(\Phi_j^*(x)\Phi_j(x)\right)dx=0, (j=1, 2, ..., N). \eqno(2.18)$$
Finally, we empasize that all the results presented here are obtained
through the analysis of the spatial part (2.1) of the Lax pair for the
BO equation. \par
\bigskip
\leftline{\bf III. LYAPUNOV STABILITY}\par
\leftline{\bf A. Variational characterization of the $N$-soliton profile}
We first show that the stationary solution $U_N$ of the higher-order BO
equation (1.9) satisfies (1.10) if one prescribes the Lagrange multipliers  $\mu_n$
appropriately.  
This provides a variational characterization of $U_N$.
Let $\Psi_j=\Phi_j^*\Phi_j$ and $b_j=-\lambda_j=a_j/2$.
With this notation,  (1.10) and (2.14) give a linear relation among $\Psi_j$
$$(N+1)\sum_{j=1}^Nb_j^{N-1}\Psi_j + \sum_{n=1}^N(-1)^{N-n+1}n\mu_n
\sum_{j=1}^Nb_j^{n-2}\Psi_j=0. \eqno(3.1)$$
In view of the fact that $\Psi_j$ are functionally independent squared eigenfunctions,$^{1,2}$
$\mu_n$ must satisfy the following system of linear algebraic equations:
$$\sum_{n=1}^N(-1)^{N-n}nb_j^{n-1}\mu_n=(N+1)b_j^{N}, (j=1, 2, ..., N). \eqno(3.2)$$
To solve (3.2), we introduce an $N\times N$ matrix $V$  
$$V=(v_{jk})_{1\leq j,k\leq N}, \ v_{jk}=b_j^{k-1}, \eqno(3.3)$$
and the cofactor of $v_{jk}$ by
$$V_{jk}={\partial |V|\over \partial v_{jk}}, \ |V|={\rm det}\ V,\eqno(3.4)$$
where $|V|$ is the Vandermonde determinant. 
Notably, since  $|V|=\prod_{1\leq j<k\leq N}(b_k-b_j)$ and  $b_j\not= b_k$ for $j\not=k$,
$|V|$ never vanishes. This fact will be used essentially in the following calculation.
It is also convenient to define
the polynomials $g(x)$ and $g_k(x)$ by
$$g(x)=\prod_{j=1}^N(x-b_j)=\sum_{s=1}^N(-1)^s\sigma_sx^{N-s}, \eqno(3.5)$$
$$g_k(x)=\prod_{j=1\atop (j\not=k)}^N(x-b_j)=\sum_{s=1}^{N-1}(-1)^s\sigma_{k,s}x^{N-s}, \eqno(3.6)$$
where $\sigma_0=1$ and $\sigma_s (1\leq s\leq N)$ are elementary symmetric functions of $b_1, b_2, ..., b_N$
$$\sigma_1=\sum_{j=1}^Nb_j, \ \sigma_2=\sum_{j,k=1\atop (j<k)}^Nb_jb_k, ...,
\ \sigma_N=\prod_{j=1}^Nb_j, \eqno(3.7a)$$
and $\sigma_{k,s}$ are given by the relation
$$\sigma_{k,s}=\sum_{j=0}^s\sigma_j(-b_k)^{s-j}. \eqno(3.7b)$$
Obviously, all $\sigma_j$ are positive quantities since $b_j>0 (j=1, 2, ..., N)$. Now, applying
Cramer's rule to (3.2) with use of the fact $|V|\not=0$, we find
that $\mu_n$ are determined uniquely as
$$\mu_n=(-1)^{N-n}{N+1\over n}{\sum_{k=1}^NV_{kn}b_k^N\over |V|},
(n=1, 2, ..., N).\eqno(3.8)$$
Substituting the formulas$^{13}$
$$V_{kn}={(-1)^{N-n}\sigma_{k,N-n}|V|\over g_k(b_k)}, \ (k, n=1, 2, ..., N),\eqno(3.9)$$ 
$$\sum_{k=1}^N{\sigma_{k,N-n}b_k^N\over g_k(b_k)}=\sigma_{N-n+1}, \ (n=1, 2, ..., N),\eqno(3.10)$$
into (3.8), we arrive at a simple expression of $\mu_n$
$$\mu_n={N+1\over n}\sigma_{N-n+1}, (n=1, 2, ..., N).\eqno(3.11)$$
If we use the relations $b_j=a_j/2 (j=1, 2, ..., N)$, we can see that $\mu_n$ are
expressed in terms of elementary symmetric functions of $a_1, a_2, ..., a_N$. \par
\leftline{\bf B. Stability}\par
Let us now prove the inequality (1.12) which assures that the functional $H_N$
is convex at the $N$-soliton profile $U_N$. 
The method used here is based on the ideas developed in a recent work$^{10}$ on
 the spectral stability
of the $N$-soliton solution of the KdV hierarchy as well as  an earlier work$^{14}$
on the algebraic structure of the BO $N$-soliton solution.
We first  rewrite (2.10) as
$$I_{n+1}(u)=(-1)^{n+1}\left\{2\pi\sum_{j=1}^Nb_j^n+(-1)^{n+1}r_n\right\}, \eqno(3.12a)$$
where we have put $b_j=-\lambda_j$ and
$$r_n={1\over 2\pi}\int^\infty_0\lambda^{n-1}\beta^*(\lambda)\beta(\lambda)d\lambda. \eqno(3.12b)$$
Let $\Delta Q$ be the increment of any functional $Q(u)$ around $u=U_N$ , i.e.,
$$\Delta Q=Q(U_N+\epsilon v)-Q(U_N). \eqno(3.13)$$
It then follows from the constraints (1.11)  that 
$$\Delta I_{n+1}=0, (n=1, 2, ..., N). \eqno(3.14)$$
We then use (3.12)  to rewrite (3.14) in the form
$$2\pi n\sum_{j=1}^Nb_j^{n-1}\Delta b_j+(-1)^{n+1}\Delta r_n=0, (n=1, 2, ..., N),\eqno(3.15)$$
where we have neglected the higher-order terms $({\Delta b_j})^s (s=2, 3, ..., N)$.
These relations indicate that the increments of  soliton amplitudes are balanced with
the increments of  radiations. We recall that $\beta\equiv 0$ for $u=U_N$ and
consequently $\Delta(\beta^*\beta)=\Delta\beta^*\Delta\beta$. This leads
to the estimates $\Delta r_n\sim O(\epsilon^2)$ and $\Delta r_n> 0$ for all $n$. 
The case $\Delta r_n=0$ calls a special attention and it will considered in detail later.
Hence, (3.15) can be solved consistently in $\Delta b_j$ only if $\Delta b_j\sim O(\epsilon^2)$. Since by the definition (1.4)
$$\Delta b_j=\epsilon\int^\infty_{-\infty}\left({\delta b_j\over\delta u}\right)_{u=U_N}v(x)dx
+O(\epsilon^2), \eqno(3.16)$$
one must impose the integral conditions on the perturbation $v(x)$
$$\int^\infty_{-\infty}\left({\delta b_j\over\delta u}\right)_{u=U_N}v(x)dx=0, \ (j=1, 2, ..., N),\eqno(3.17)$$
in accordance with the above estimate for $\Delta b_j$.
We can see from (2.12), (2.14) together with the relations $b_j=-\lambda_j (j=1, 2, ..., N)$ and $|V|\not=0$ that (3.17) are  equivalent to
$$\int^\infty_{-\infty}\left({\delta I_{n+1}\over\delta u}\right)_{u=U_N}v(x)dx=0, \ (n= 1, 2, ..., N).\eqno(3.18)$$
Owing to (3.14), however, these conditions are satisfied automatically.  The above observations allow us to solve
(3.15). Indeed, the solutions are written, with use of Cramer's rule, as
$$\Delta b_j={1\over 2\pi}{\sum_{n=1}^N{(-1)^n\over n}V_{jn}\Delta r_n\over |V|}. \ (j=1, 2, ..., N).\eqno(3.19)$$
It now follows from (1.9b), (3.12) and (3.14) that
$$\Delta H_N=(-1)^N\left\{2\pi (N+1)\sum_{j=1}^Nb_j^N\Delta b_j+(-1)^{N+2}\Delta r_{N+1}\right\}. \eqno(3.20)$$
If we substitute (3.19) into (3.20) and use the fomulas (3.9) and (3.10), $\Delta H_N$ simplifies to
$$\Delta H_N=(N+1)\sum_{n=1}^N{\sigma_{N-n+1}\over n}\Delta r_n+\Delta r_{N+1}. \eqno(3.21)$$
Since $\sigma_{N-n+1}>0$ for $n=1, 2, ..., N$ by the definition (3.7),
we find that if at least one of $\Delta r_n$ is not zero, then $\Delta H_N>0$.
On the other hand, if all $\Delta r_n$ become zero, then $\Delta H_N=0$. In the 
latter case, we see from (3.15) that $\Delta b_j=0$ for all $j$. This situation will 
happen when the perturbation $\epsilon v$ represents the small variation of $U_N$
with respect to the phase parameters $x_{j0}$. Specifically
$$\epsilon v(x)=\sum_{j=1}^N{\partial U_N\over\partial x_{j0}}\delta x_{j0}, \eqno(3.22)$$
where $\delta x_{j0}$ are small perturbations of order $\epsilon$.
If we impose the following $N$ integral conditions on $v(x)$ in addition to (3.28)
$$\int^\infty_{-\infty}{\partial\over\partial x}\left({\delta I_{n+1}\over \delta u}\right)_{u=U_N}
v(x)dx =0, (n=1, 2, ..., N), \eqno(3.23)$$
then the perturbation of the form (3.22) ceases to be admissible, as we shall now demonstrate.
We first notice that the right-hand side of (3.22) can be expressed in terms of the $x$ derivative
of $(\delta I_{n+1}/\delta u)_{u=U_N}$. Indeed, 
we take $u=u_N$ in (1.3) and then put
$t_0=t_1=...=0$  to obtain
$$(-1)^{n}n\sum_{j=1}^Nb_j^{n-1}\ {\partial U_N\over\partial x_{j0}}
={\partial\over\partial x}\left({\delta I_{n+1}\over \delta u }\right)_{u=U_N},
(n=1, 2. ..., N), \eqno(3.24)$$
where we have used (1.6), the definition of $U_N$ and $b_j=a_j/2$.
Thanks to the fact $|V|\not=0$, the relations (3.24)
can be inverted to give
$${\partial U_N\over\partial x_{j0}}=\sum_{n=1}^{N}{(-1)^n\over n}{V_{jn}\over |V|}
{\partial\over\partial x}\left({\delta I_{n+1}\over \delta u}\right)_{u=U_N},
 (j=1, 2, ..., N). \eqno(3.25)$$
An alternative expression of (3.22) follows immediately upon introducing (3.25) into (3.22), which reads
$$\epsilon v(x)=\sum_{j=1}^N\sum_{n=1}^{N}{(-1)^n\over n}{V_{jn}\over |V|}
{\partial\over\partial x}\left({\delta I_{n+1}\over \delta u}\right)_{u=U_N}\delta x_{j0}.
\eqno(3.26)$$
 We observe that this perturbation satisfies the conditions (3.18) by virtue of (2.11).
 It is important that the $N\times N$ matrix $C=(c_{jk})_{1\leq j,k\leq N}$ with elments
 $$c_{jk}=\int^\infty_{-\infty}{\partial\over\partial x}\left({\delta I_{j+1}\over \delta u}\right)_{u=U_N}
 {\partial\over\partial x}\left({\delta I_{k+1}\over \delta u}\right)_{u=U_N}dx, \eqno(3.27)$$
 is positive definite and hence $|C|\not= 0$.
In view of this fact, we deduce from (3.23) and (3.26) that $\delta x_{j0}=0 (j=1, 2, ..., N)$ and consequently
$v=0$, which 
implies the assertion mentioned above. An additional relation which deserves remark is  
$$\int^\infty_{-\infty}\left({\delta\beta\over\delta u}\right)_{u=U_N} v(x)dx=0, \eqno(3.28)$$
which follows from (2.14), (2.17), (2.18) and (3.26).
This  leads to the estimates $\Delta\beta\sim O(\epsilon^2)$ and 
$\Delta r_n\sim O(\epsilon^4)$. As a result, the pertubation (3.26) gives rise to higher-order
contributions to (3.21) which also means that the second variation of $H_N$ turns out to be zero.
In conclusion,  the inequality $\Delta H_N>0$  holds 
under the simultaneous conditions
(3.18) and (3.25) imposed on $v(x)$,  which completes the proof of
 (1.12). 
The convexity of $H_N$ implies that the second variation of $H_N$ is strictly positive
and consequently the $N$-soliton solutions  are  energetically stable.
\par
\bigskip
\leftline{\bf C. Remark}\par
In this paper, the convexity of $H_N$ has been proved by  invoking some
results obtained by the IST of the BO equation. There exists, however 
another method to establish the same convex property without recourse to the IST.
To illustrate this, we put $u(x,t)=U_N(x)+\epsilon v(x)e^{\lambda t}$ and linearize
 (1.9) around $U_N$. The resulting eigenvalue equation can be written as
 $${\partial\over\partial x}{\cal L}_Nv=\lambda v, \eqno(3.29)$$
 where ${\cal L}_N$ is a self-adjoint operator. This operator may be
 defined through the relation
 $$\delta^2H_N={\epsilon^2\over 2}\int^\infty_{-\infty}v(x){\cal L}_Nv(x)dx, 
\eqno(3.30)$$
 where $\delta^2H_N$ denotes the second variation of $H_N$. 
Let $n({\cal L}_N)$ be the number of negative eigenvalues of ${\cal L}_N$ and
$p({\cal H}_N)$  be the number of positive eigenvalues of the Hessian matrix
defined by
$${\cal H}_N=(h_{jk})_{1\leq j,k\leq N}, 
\ h_{jk}={\partial^2 H_N\over \partial \mu_j\partial \mu_k}. \eqno(3.31)$$
Then, under the conditions (3.18) and (3.23) the positivity of $\delta^2 H_N$
is satified if and only if $n({\cal L}_N)=p({\cal H}_N)$.
The above criterion of the positivity property has been proved in Ref. 15 and
has been applied to the Lyapunov stability of the $N$-soliton solution of
the KdV equation.$^{9}$  In particular, the spectral property of the $2N$th-order
differential operator associated withe the linearized KdV equation 
 has been investigated by extending the classical
Sturmian theory. See also a related work dealing with the stability of the
$N$-soliton solutions in the KdV hierarchy.$^{10}$
In the case of the BO equation, however  the
eigenvalue equation   (3.29) is
 not  purely differential equation but actually 
 integrodifferential equation
  since it includes the Hilbert transform.
 This makes the spectral analysis more difficult. Quite recentry, a new method
 was developed to characterize the spectral property of  ${\cal L}_N$ for
 $N=2$.$^{8}$ The extension to the general $N$-soliton solutions of the BO equation and its 
 hierarchy is still  to be resolved.  It is noteworthy that   $ p({\cal H}_N)$
 can be evaluated explicitly for the $N$-soliton profile $U_N$. This calculation is
 presented in Appendix B. The  stability analysis developed in this paper 
 would suggest that $n({\cal L}_N)$
  is equal to $ p({\cal H}_N)$.  This interesting issue will be
  persued in a future study. \par
 \bigskip
 \leftline{\bf APPENDIX A: PROOF OF THE $N$-SOLITON SOLUTION}\par
In this Appendix, we provide a direct proof of the $N$-soliton solution (1.7) of
the $n$th higher-order BO equation (1.3)  by means of
an elementary theory of determinants. For convenience, we write down
some basic formulas for determinants upon which our proof relies.
Let $F$ be an $N\times N$ matrix with elements $f_{jk}$ given by (1.7b) and 
$F_{jk}$ be the cofactor of  $f_{jk}$  The expansion of $|F|$
by elements and their cofactors is given by the  two ways:
$$\sum_{k=1}^Nf_{jk}F_{lk}=\delta_{jl}|F|, \eqno(A1a)$$
$$\sum_{j=1}^Nf_{jk}F_{jl}=\delta_{kl}|F|. \eqno(A1b)$$
The folowing formula is a consequence of (A1)
$$\sum_{j,k=1}^N(f_j+g_k)f_{jk}F_{jk}=\sum_{j=1}^N(f_j+g_j)|F|. \eqno(A2)$$
The differential rule applied to the determinant $|F|$ gives
$$|F|_x=\sum_{j=1}^NF_{jj}, \eqno(A3a)$$
$$|F|_{t_n}=(-1)^n\sum_{j=1}^Nc_jF_{jj}, \eqno(A3b)$$
where $c_j= (n+1)a_j^{n}/2^n$.
To carry out the proof, it is necessary to assign the time dependence of
the eigenfunction $\Phi_j$ for the discrete spectrum $\lambda_j$. This
can be accomplished simply by replacing the phase factor $\gamma_j$
introduced in (2.13) by $x_j$ which is defined in (1.6). We first
show that (2.15) can be rewritten in an alternative determinantal form (1.7).
The solution to  (2.13) is found by using Cramer's rule as
$$\Phi_j=i\sum_{k=1}^N{F_{kj}\over |F|}. \eqno(A4)$$
We put $f_j=a_j$ and $g_j=-a_j $ in (A2) to derive the relation
$$\sum_{j,k=1}^NF_{jk}=\sum_{j=1}^NF_{jj}. \eqno(A5)$$
It follows from (A3)-(A5) that
$$\sum_{j=1}^N\Phi_j=i({\rm ln}\ |F|)_x. \eqno(A6)$$
Substituting  (A6) and its complex conjugate expression into
(2.15), we find that (2.15) coincides with (1.7). \par
Let us now proceed to the proof of the
$N$-soliton solution. We substitute (1.7) and (2.14) into (1.3) and
integrate it once with respect to $x$ to recast (1.3) into the form
$$i(|F|^*|F|_{t_n}-|F||F|^*_{t_n})/|F|^*|F|
=(-1)^n(n+1)\sum_{j=1}^N\left({a_j\over 2}\right)^{n-1}\Phi_j^*\Phi_j, \eqno(A7)$$
where we have used the relation $\lambda_j=-a_j/2$.
The following identity has been established by using Jacobi's formula for determinants:
$$i\left({|F|^*F_{jk}\over a_k}-{|F|F_{kj}^*\over a_j}\right)
=2{(|F|\Phi_j)^*(|F|\Phi_k)\over  a_ja_k}, (j, k=1, 2, ..., n). \eqno(A8)$$
Indeed,  (A8) coinsides with  (A20) in Ref. 7  
with the identification $f=|F|^*, \Delta_{jk}=F_{jk}^*, \psi_j=\Phi_j^*$.
If we multiply (A8) with $j=k$ by $c_j$ and sum up with respect to $j$, 
we obtain
$${i\over 2}\sum_{j=1}^Nc_j(|F|^*F_{jj}-|F|F_{jj}^*)
={n+1\over 2^n}\sum_{j=1}^Na_j^{n-1}\Phi_j^*\Phi_j|F|^*|F|. \eqno(A9)$$
The  left-hand side of (A9) is modified further by introducing the formula (A3b)
and its complex conjugate expression. It leads, after dividing the resultant
expressin by $|F|^*|F|$, to (A7) and thus completing the proof. \par
\bigskip
\leftline{\bf APPENDIX B: POSITIVE EIGENVALUES OF THE HESSIAN MATRIX ${\cal H}_N$}\par
The Hessian matrix ${\cal H}_N$ is defined by (3.31). It is a real symmetric matrix 
whose elements are calculated explicitly for the $N$-soliton solution. 
Indeed, by taking $\beta=0$ in (2.10),  the $n$th conservation law  corresponding to $u=u_N$ reduces to
$$I_n=2\pi (-1)^n\sum_{l=1}^Nb_l^{n-1}, \ (b_l=-\lambda_l). \eqno(B1)$$
If we regard $H_N$ as a function of $\mu_j (j=1, 2, ..., N)$, we obtain from (1.9b) and (1.10) $${\partial H_N\over\partial \mu_j}=I_{j+1}, (j=1, 2, ..., N). \eqno(B2)$$
Hence
$$h_{jk}={\partial I_{j+1}\over\partial \mu_k}
=2\pi (-1)^{j+1}j\sum_{l=1}^Nb_l^{j-1}{\partial b_l\over\partial \mu_k}.\eqno(B3)$$
Let $P=(p_{jk})_{1\leq j,k\leq N}$ and $Q=(q_{jk})_{1\leq j,k\leq N}$ be $N\times N$ matrices with
elements
$$p_{jk}=2\pi (-1)^{j+1}jb_k^{j-1}, \eqno(B4a)$$
$$q_{jk}={\partial \mu_j\over \partial b_k}={N+1\over j}{\partial \sigma_{N-j+1}\over \partial b_k},\eqno(B4b)$$
respectively. Note that the right-hand side of (B4b) follows from (3.11). Using the above definition,
we can rewrite (B3) in the form
$${\cal H}_N=PQ^{-1},\eqno(B5)$$
if $Q^{-1}$ exists. To show the nonsingular nature of $Q$, we use the definition (3.7a) of $\sigma_{N-j+1}$ and (B4b)
to evaluate the determinant of $Q$. A simple calculation immediately leads to
$$|Q|={(N+1)^N\over N!}\prod_{1\leq j<k\leq N}(b_k-b_j).\eqno(B6)$$
Since $b_j\not=b_k$ for $j\not=k$, we confirm that $|Q|\not=0$, implying that $Q$ is invertible. \par
It now follows from (B5) that
$$Q^T{\cal H}_NQ=Q^TP.\eqno(B7)$$
In accordance with Sylvester's law of inertia, one can see from (B7) that the number of
positive eigenvalues of ${\cal H}_N$ coincides with that of $Q^TP$. The latter
can be counted easily, as we shall now demonstrate.
Using (B4), the $(j, k)$ element of $Q^TP$ becomes
$$(Q^TP)_{jk}=2\pi(N+1)\sum_{l=1}^N(-1)^{l+1}{\partial\sigma_{N-l+1}\over\partial b_j}b_k^{l-1}. \eqno(B8)$$
We differentiate (3.5) by $b_j$ and then put $x=b_k$ to derive the relation
$$\sum_{l=1}^N(-1)^{l+1}{\partial\sigma_{N-l+1}\over\partial b_j}b_k^{l-1}
=\delta_{jk}\prod_{l=1\atop (l\not=j)}(b_l-b_k).\eqno(B9)$$
Introducing (B9) into (B8), we find that
 $Q^TP$ is a diagonal matrix. We can order the magnitude of
$b_j$ as $b_1>b_2>...>b_N>0$ without loss of generality. Then, (B8) and (B9) 
indicate that the number of positive eigenvalues of $Q^TP$ is equal to
$\left[{N+1\over 2}\right]$ where $[x]$ denotes the integer part of $x$.
If we take account of (B7) and Sylvester's law of inertia, we conclude
that $p[{\cal H}_N]$ =$\left[{N+1\over 2}\right]$. \par
\bigskip

\newpage
\leftline{\bf REFERENCES}\par
\begin{enumerate}
\item D.J. Kaup, T.I. Lakoba and Y. Matsuno, Phys. Lett. A {\bf 238}, 123 (1998).
\item D.J. Kaup, T.I. Lakoba and Y. Matsuno, Inverse Problems  {\bf 15},  215 (1999).
\item Y. Matsuno, J. Phys. Soc. Jpn. {\bf 47}, 1745 (1979).
\item H.H. Chen and D.J. Kaup, Phys. Fluids  {\bf 23}, 235 (1980).
\item D.P. Bennett, R.W. Brown, S.E. Stansfield,  J.D. Stroughair and J.L. Bona,
       Math. Proc. Camb. Phil. Soc. {\bf 94}, 351 (1983).
\item Y. Matsuno and D. J. Kaup, Phys. Lett. A {\bf 228},  176 (1997).
\item Y. Matsuno and D. J. Kaup, J. Math. Phys. {\bf  38}, 5198 (1997).
\item A. Neves and O. Lopes, Comm. Math. Phys. {\bf 262}, 757 (2006).
\item J. Maddocks and R. Sachs, Comm. Pure Appl. Math. {\bf 46}, 867 (1993).
\item Y. Kodama and D. Pelinovsky, J. Phys. A: Math. Gen. {\bf 38}, 6129 (2005).
\item A.S.  Fokas and  M.J.  Ablowitz, Stud. Appl. Math. {\bf 68}, 1 (1983).
\item D.J. Kaup and Y. Matsuno,  Stud. Appl. Math. {\bf 101}, 73 (1998).
\item R. Vein and P. Dale, {\it Determinants and Their Applications in
      Mathematical Physics} (Springer, New York, 1999).
\item Y. Matsuno, J. Phys. Soc. Jpn. {\bf 51}, 3375 (1982).
\item M. Grillakis, J. Shatah and W. Strauss, J. Func. Anal. {\bf 94}, 308 (1990).
\end{enumerate}

\end{document}